\documentclass[runningheads]{llncs}
\usepackage{graphicx}
\begin{document}
\title{Automated segmentation of intracranial hemorrhages from 3D CT \protect\\ INSTANCE 2022 challenge report}
%
%
\author{Md Mahfuzur Rahman Siddiquee \and
Dong Yang \and
Yufan He \and
Daguang Xu \and
Andriy Myronenko
}

\authorrunning{M. Rahman Siddiquee et al.}
%
\institute{NVIDIA, Santa Clara, CA \\
\email{\{mdmahfuzurr,dongy,yufanh,daguangx,amyronenko\}@nvidia.com}}
\maketitle             

\begin{abstract}

Intracranial hemorrhage segmentation challenge (INSTANCE 2022) offers a platform for researchers to compare their solutions to segmentation of hemorrhage stroke regions from 3D CTs. In this work, we describe our solution  to INSTANCE 2022. We use a 2D segmentation network, SegResNet from MONAI, operating slice-wise without resampling.  The final submission is an ensemble of 18 models.  Our solution (team name NVAUTO) achieves the top place in terms of Dice metric (0.721), and overall rank 2 (based on the combined metric ranking\footnote{https://instance.grand-challenge.org/}). It is implemented  with Auto3DSeg\footnote{https://monai.io/apps/auto3dseg}.

\end{abstract}

\section{Introduction}

\begin{figure}[!ht]
    \centering
    \includegraphics[width=0.32\textwidth]{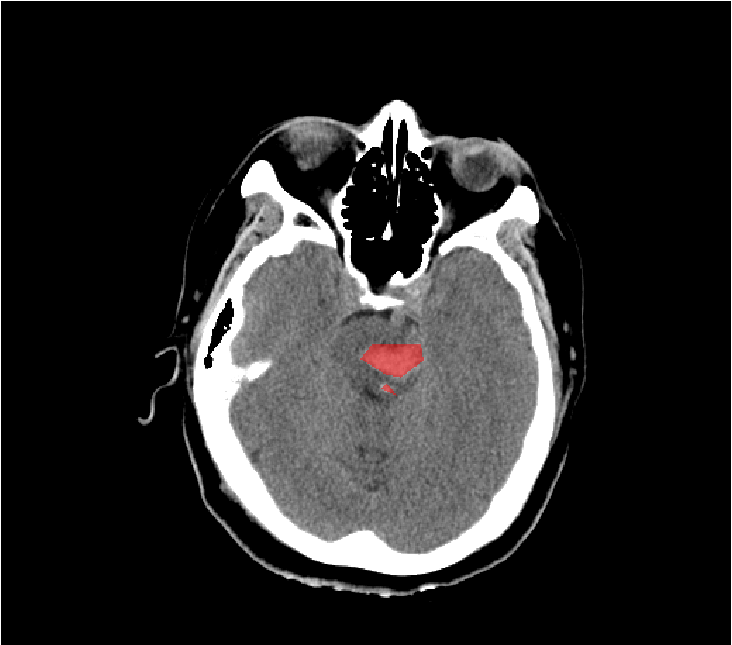}
    \includegraphics[width=0.32\textwidth]{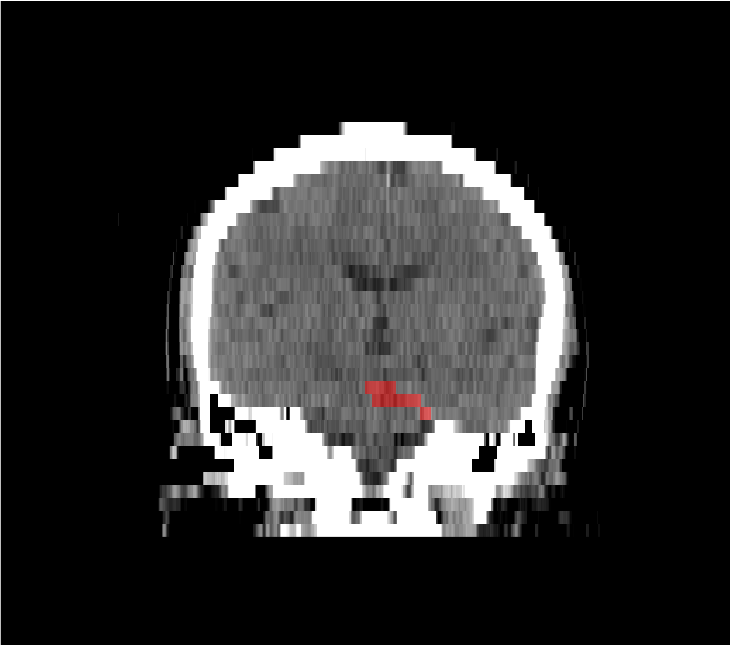}
    \includegraphics[width=0.32\textwidth]{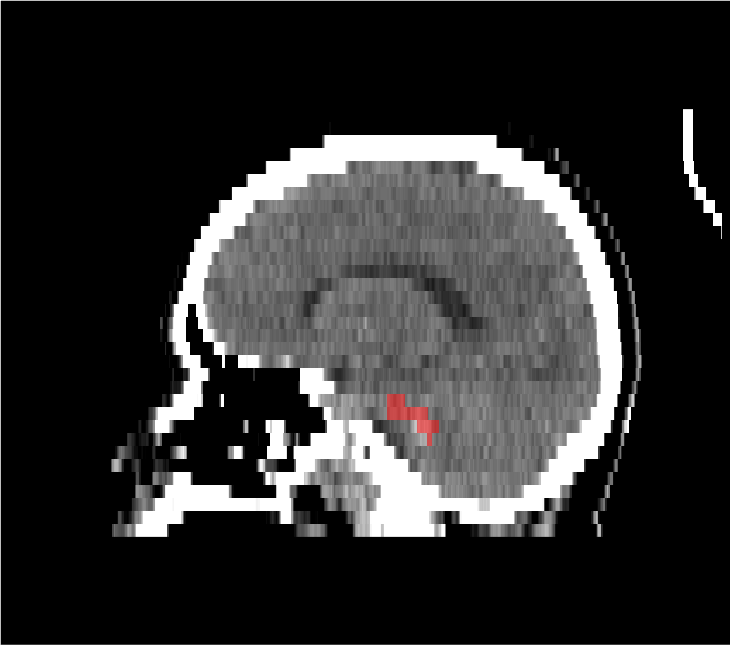}
    \caption{An example case with 3D CT image and the annotated hemorrhage stroke region.}
    \label{fig:example1}
\end{figure}

Segmentation of intracranial hemorrhage (ICH) is necessary for early and accurate patient diagnosis and can help save patients lives during early onset. ICH is a common stroke type and has the highest mortality rate among all stroke types~\cite{instance2}. Intracranial Hemorrhage Segmentation challenge (INSTANCE) 2022 aims to advance novel automatic segmentation methods for accurate intracranial hemorrhage segmentations on non-contrast 3D CT images~\cite{instance1}. The INSTANCE22 dataset consists of 200 cases (100 labeled cases were provided for training) . Each case includes a 3D CT with refined labeling from 10 experienced radiologists. The task is to segment a single class (stroke region).  An example case with 3D CT and the corresponding ground-truth stroke region overlays is shown in Figure~\ref{fig:example1}.

\section{Method}

\begin{figure}[t]
    \centering
    \includegraphics[width=0.8\textwidth]{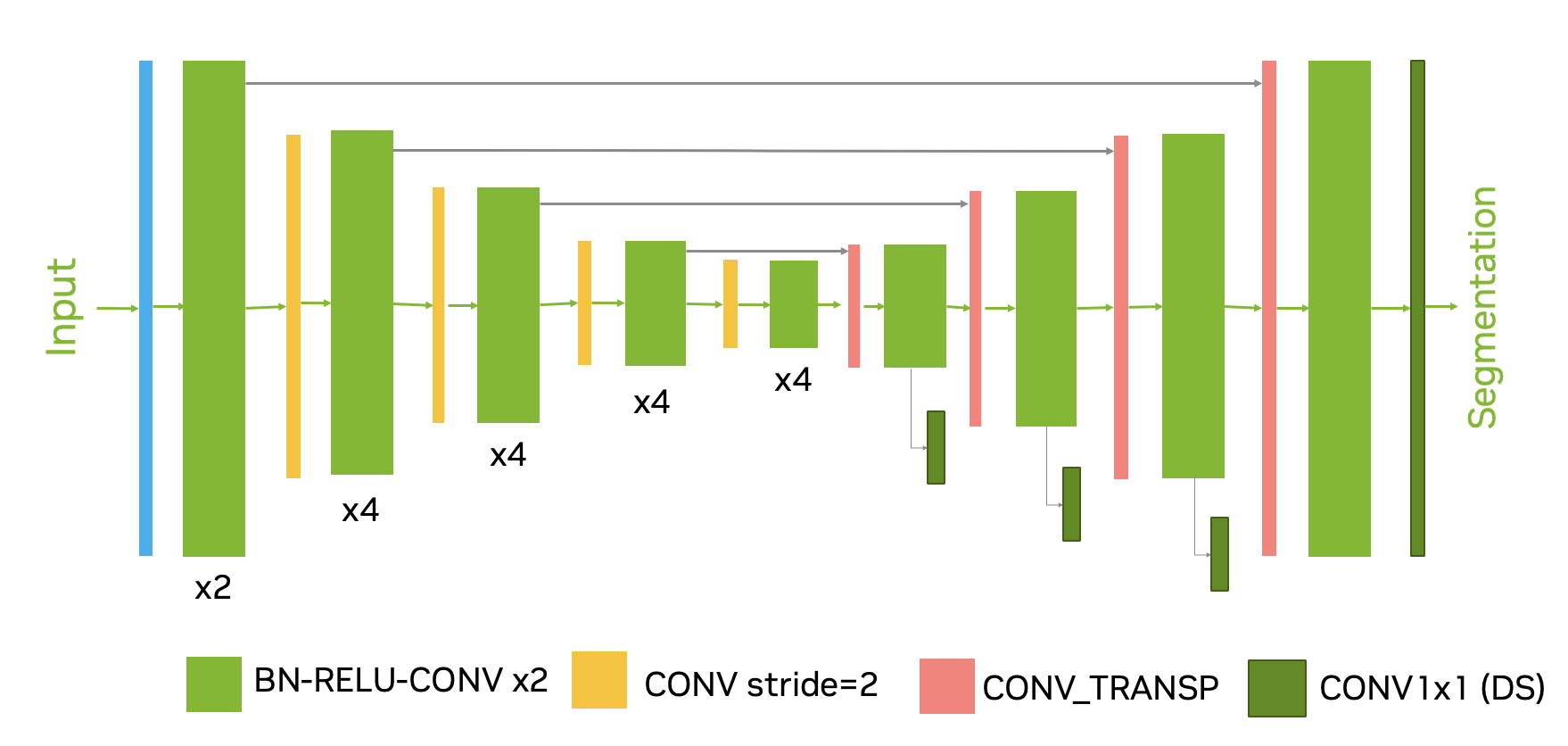}
    \caption{SegResNet 2D network configuration. The network uses repeated ResNet blocks with batch normalization and deep supervision}
    \label{fig:net}
\end{figure}

We implemented our approach with MONAI\footnote{https://github.com/Project-MONAI/MONAI}~\cite{monai}, we used Auto3Dseg\footnote{https://monai.io/apps/auto3dseg} system to automate most parameter choices. We decided to use a 2D segmentation and operate slice-by-slice, mainly because in this dataset a typical in-plane resolution is 10x higher than out-of-plane (about 0.46x0.46x5mm), and there were a minimal continuity of segmentation masks between slices (some cases had only 2-3 slices labeled). For the main network architecture we used SegResNet\footnote{https://docs.monai.io/en/stable/networks.html\#segresnet}, which is an encode-decoder based semantic segmentation network based on~\cite{myronenko20183d}, with deep supervision (see Figure~\ref{fig:net}).

The encoder part uses ResNet~\cite{he2016identity} blocks with instance normalization. We have used 5 stages of down-sampling, each stage has 2, 4, 4, 4, and 4 convolutional blocks, respectively. Each block's output is followed by an additive identity skip connection. We follow a common CNN approach to downsize image dimensions by 2 progressively and simultaneously increase feature size by 2.  All convolutions are 3x3 with an initial number of filters equal to 32. The encoder is trained with $384\times384$ input region. The decoder structure is similar to the encoder one, but with a single block per each spatial level. 
 The end of the decoder has the same spatial size as the original image, and the number of features equal to the initial input feature size, followed by a 1x1x1 convolution and a softmax. For the deep supervision of 3 sub-levels of the decoder branch features, we add extra projections heads (1x1 convolutions) into the same number of output classes. These additional network outputs are used to compute  losses at smaller image scales. 

\subsubsection{Dataset} We use the INSTNACE22 dataset~\cite{instance1}  for training the model.  We randomly split the entire dataset into 5 folds and trained a model for each fold. 

\subsubsection{Data normalization and input} We use three  strategies for 3D CT image normalization and input:
\begin{itemize}
    \item 3 adjacent 2D slices at a single CT window/level (as a 3 channels 2D input)
    \item one 2D slice at 3 different CT window/levels (as a 3 channels 2D input)
    \item combination of the above (as a 9 channels 2D input)
\end{itemize}
all these strategies had similar validation performance, we use the ensemble of all of them for the submission. 

\subsubsection{Cropping} We crop a random patch of 384x384 pixels, which includes most of the foreground content. We use the batch size of 32.

\subsubsection{Augmentations} We use random a) flips (all axes)
b) random rotation and scaling,  c) noise, intensity scale/shift, patch shuffle 

\subsubsection{Loss} We use the combined Dice + CrossEntropy loss. The same loss is summed over all deep-supervision sublevels:

\begin{equation}
Loss= \sum_{i=0}^{4} \frac{1}{2^{i}} Loss(pred,target^{\downarrow}) 
\end{equation}
where the weight $\frac{1}{2^{i}}$ is smaller for each sublevel (smaller image size) $i$. The target labels are downsized (if necessary) to match the corresponding output size using nearest neighbor interpolation. 

\subsubsection{Optimization} We use the AdamW optimizer with an initial learning rate of $2e^{-4}$ and decrease it to zero at the end of the final epoch using the Cosine annealing scheduler. All the models were trained for 1600 epochs with deep supervision. We use batch size of 32 per GPU, and train on 8 GPUs 16Gb NVIDIA V100 DGX machine (which is equivalent to batch size of 256). We use weight decay regularization of $1e^{-5}$.

\section{Results}

Based on our data splits, a single run 5-folds cross-validation results are shown in Table~\ref{tab:result}.  On average, we achieve $0.7959$ cross-validation performance in terms of  Dice metric.

\begin{table}[h!]
    \centering
    \begin{tabular}{| c | c | c | c | c | c |}
        \hline
        {\textbf{Fold 1}} & {\textbf{Fold 2}} & {\textbf{Fold 3}} & {\textbf{Fold 4}} & {\textbf{Fold 5}} & {\textbf{Average}} \\
        \hline
         0.7303	 & 0.8513 &	0.8205	& 0.7619 &	0.8155 & 0.7959 \\
        \hline
    \end{tabular}
    \caption{Dice accuracy using 5-fold cross-validation.}
    \label{tab:result}
\end{table}

For the final submission we use a mean ensemble of 18 models total (3 fully trained runs, using 5-folds best checkpoints, and a few extra checkpoints). Table~\ref{tab:result2} shows the final ranking and scores on the hidden test sets (provided by the organizers) for the top 3 places. Our solution (NVAUTO) team achieves the best Dice accuracy, but ranks 2nd overall based on all 4 metrics used in this challenge (Dice accuracy, Relative volume difference (RD), Normalized surface dice (NSD), Hausdorff distance (HD)). 
 
\begin{table}[h!]
    \centering
    \begin{tabular}{| l | l | c | c | c | c |}
        \hline
        \textbf{Rank} & \textbf{Team} & \textbf{Dice} & \textbf{RVD} & \textbf{NSD} &  \textbf{HD}  \\
        \hline
        1 & Vegetable~\cite{team_vegetable22} & 0.693 & 0.212 & 0.536 & 35.272\\
        \hline
        2 & NVAUTO (ours) & 0.721 & 0.261 & 0.534 & inf \\
        \hline
        3 & MEC-LAB~\cite{team_meclab22} & 0.690 & 0.313 & 0.513 & inf\\
        \hline
    \end{tabular}
    \caption{Top 3 teams of the INSTANCE22 challenge with the corresponding 4 metrics used for the final ranking: Dice accuracy, Relative volume difference (RD), Normalized surface dice (NSD), Hausdorff distance (HD).}
    \label{tab:result2}
\end{table}

\bibliographystyle{splncs04}
\bibliography{paper}

\begin{thebibliography}{1}
\providecommand{\url}[1]{\texttt{#1}}
\providecommand{\urlprefix}{URL }
\providecommand{\doi}[1]{https://doi.org/#1}

\bibitem{monai}
Project-monai/monai, \url{https://doi.org/10.5281/zenodo.5083813}

\bibitem{he2016identity}
He, K., Zhang, X., Ren, S., Sun, J.: Identity mappings in deep residual
  networks. In: European conference on computer vision. pp. 630--645. Springer
  (2016)

\bibitem{team_vegetable22}
Li, C., Chen, Z.: {nnU-Net} for intracranial hemorrhage segmentation. In:
  {INSTANCE22 report, MICCAI2022}. Southern Medical University, Gaungzhou,
  China (2022)

\bibitem{instance2}
Li, X., Luo, G., Wang, W., Wang, K., Gao, Y., Li, S.: Hematoma expansion
  context guided intracranial hemorrhage segmentation and uncertainty
  estimation. IEEE Journal of Biomedical and Health Informatics
  \textbf{26}(3),  1140--1151 (2022). \doi{10.1109/JBHI.2021.3103850}

\bibitem{instance1}
Li, X., Wang, K., Liu, J., Wang, H., Xu, M., Liang, X.: {The 2022 Intracranial
  Hemorrhage Segmentation Challenge on Non-Contrast head CT (NCCT)} (Mar 2022).
  \doi{10.5281/zenodo.6362221}, \url{https://doi.org/10.5281/zenodo.6362221}

\bibitem{myronenko20183d}
Myronenko, A.: {3D MRI} brain tumor segmentation using autoencoder
  regularization. In: International MICCAI Brainlesion Workshop. pp. 311--320.
  Springer (2018)

\bibitem{team_meclab22}
Saner, A.: Perimeter-based losses evaluation for intracranial hemorrhage
  segmentation. In: {INSTANCE22 report, MICCAI2022}. Technical University of
  Darmstadt, Germany (2022)

\end{thebibliography}
\end{document}